\documentclass[twocolumn]{aastex62}
\usepackage{csvsimple}
\usepackage{multirow}
\usepackage{amsmath}
\usepackage{longtable}
\usepackage{booktabs}
\usepackage{ltablex}
\usepackage{epsfig}
\usepackage{threeparttable}

\def\cm2{\rm \ cm$^{-2}$}
\def\ss{\mbox{s\,s$^{-1}$}}

\def\gleam{\mbox{GLEAM-X\,J1627\ensuremath{-}52}}
\def\gpm{\mbox{GPM\,J1839\ensuremath{-}10}}
\def\mtp{\mbox{PSR\,J0901\ensuremath{-}4046}}

\shorttitle{Population study of long-period radio transients in the neutron-star and white-dwarf scenarios}
\shortauthors{Rea, Hurley-Walker, Pardo-Araujo et al.}

\begin{document}

\title{Long-period radio pulsars: population study in the neutron star and white dwarf rotating dipole scenarios}

\correspondingauthor{Nanda Rea}
\email{rea@ice.csic.es}

\author[0000-0003-2177-6388]{N. Rea} 
\affiliation{Institute of Space Sciences (ICE), CSIC, Campus UAB, Carrer de Can Magrans s/n, E-08193, Barcelona, Spain}
\affiliation{Institut d'Estudis Espacials de Catalunya (IEEC), Carrer Gran Capit\`a 2--4, E-08034 Barcelona, Spain}

\author{N. Hurley-Walker}
\affiliation{International Centre for Radio Astronomy Research, Curtin University, Kent St, Bentley WA 6102, Australia}

\author{C. Pardo-Araujo} 
\affiliation{Institute of Space Sciences (ICE), CSIC, Campus UAB, Carrer de Can Magrans s/n, E-08193, Barcelona, Spain}
\affiliation{Institut d'Estudis Espacials de Catalunya (IEEC), Carrer Gran Capit\`a 2--4, E-08034 Barcelona, Spain}

\author{M. Ronchi} 
\affiliation{Institute of Space Sciences (ICE), CSIC, Campus UAB, Carrer de Can Magrans s/n, E-08193, Barcelona, Spain}
\affiliation{Institut d'Estudis Espacials de Catalunya (IEEC), Carrer Gran Capit\`a 2--4, E-08034 Barcelona, Spain}

\author{V. Graber} 
\affiliation{Institute of Space Sciences (ICE), CSIC, Campus UAB, Carrer de Can Magrans s/n, E-08193, Barcelona, Spain}
\affiliation{Institut d'Estudis Espacials de Catalunya (IEEC), Carrer Gran Capit\`a 2--4, E-08034 Barcelona, Spain}

\author{F. Coti Zelati} 
\affiliation{Institute of Space Sciences (ICE), CSIC, Campus UAB, Carrer de Can Magrans s/n, E-08193, Barcelona, Spain}
\affiliation{Institut d'Estudis Espacials de Catalunya (IEEC), Carrer Gran Capit\`a 2--4, E-08034 Barcelona, Spain}

\author{D.~de Martino}
\affiliation{INAF – Capodimonte Astronomical Observatory Naples Via Moiariello 16, I-80131 Naples, Italy}

\author{A.~Bahramian}
\affiliation{International Centre for Radio Astronomy Research, Curtin University, Kent St, Bentley WA 6102, Australia}

\author{S.~J.~McSweeney}
\affiliation{International Centre for Radio Astronomy Research, Curtin University, Kent St, Bentley WA 6102, Australia}

\author{T.~J.~Galvin}
\affiliation{CSIRO, Space and Astronomy, PO Box 1130, Bentley WA 6102, Australia}
\affiliation{International Centre for Radio Astronomy Research, Curtin University, Kent St, Bentley WA 6102, Australia}

\author{S.~D.~Hyman}
\affiliation{Department of Engineering and Physics, Sweet Briar College, Sweet Briar, VA 24595, USA}

\author{M.~Dall'Ora}
\affiliation{INAF – Capodimonte Astronomical Observatory Naples Via Moiariello 16, I-80131 Naples, Italy}


\received{July, 2023}
\submitjournal{ApJ}

\begin{abstract}
The nature of two recently discovered radio emitters with unusually long periods of 18\,min (\gleam) and 21\,min (\gpm) is highly debated. Their bright radio emission resembles that of radio magnetars, but their long periodicities and lack of detection at other wavelengths challenge the neutron-star interpretation. In contrast, long rotational periods are common in white dwarfs but, although predicted, dipolar radio emission from isolated magnetic white dwarfs has never been unambiguously observed. In this work, we investigate these long-period objects as potential isolated neutron-star or white-dwarf dipolar radio emitters and find that both scenarios pose significant challenges to our understanding of radio emission via pair production in dipolar magnetospheres. We also perform population-synthesis simulations based on dipolar spin-down in both pictures, assuming different initial-period distributions, masses, radii, beaming fractions, and magnetic-field prescriptions, to assess their impact on the ultra-long pulsar population. In the neutron-star scenario, we do not expect a large number of ultra-long period pulsars under any physically motivated (or even extreme) assumptions for the period evolution. 
On the other hand, in the white-dwarf scenario, we can easily accommodate a large population of long-period radio emitters. However, no mechanism can easily explain the production of such bright coherent radio emission in either scenarios.
\end{abstract}



\section{Introduction} \label{sec:intro}

Highly polarized and periodic Galactic radio sources have been historically interpreted as rotating magnetic neutron-star (NS) dipoles. 
Until recently, measured periods clustered between about 1\,ms -- 20\,s in line with predictions of recycling scenarios \citep{B&H1991,Tauris2012} for the fast and magnetic-field decay \citep{2013MNRAS.434..123V, Pons2013} for the slow rotators. The discovery of ultra-long periodic coherent radio emitters challenges these models. Some seem to be extreme NS pulsars (e.g., the 76-s source \mtp; \citealt{2022NatAs...6..828C}), while the interpretation of others is still uncertain (e.g., the 18-minute source \gleam; \citealt{2022Natur.601..526H}). The periodic radio emission of \gleam\, lasted only a few months in 2018 with a flux density (20-50\,Jy) and polarization degree ($\sim$90\% linear) similar to other radio magnetars, and its long periodicity can be explained through post-supernova fallback accretion \citep[see, e.g.,][]{Alpar2001, Chatterjee2000, Ertan2009, Tong2016, 2022ApJ...934..184R}. 
However, deep X-ray limits challenge the magnetar interpretation \citep{2022ApJ...940...72R}.

In contrast, slow spin periods are common in magnetic white dwarfs (MWDs) \citep{Ferrario2005,Ferrario2020}. Although isolated MWDs exhibit magnetic-field strengths between $10^6$ to $10^9$\,G \citep{Ferrario2015, Ferrario2020}, lower than NS $B$-fields spanning $10^8$ to $10^{15}$\,G, MWD have also been proposed to emit spin-down-driven radio emission similar to NSs \citep{Zhang2005}. To date, two radio-emitting WDs have been detected, the binary systems AR\,Sco  ($P\sim1.95$\,min in a 3.5\,hr orbit; \citealt{2016Natur.537..374M}) and J1912$-$4410 ($P\sim 5.3$\,min in a 4\,hr orbit; \citealt{Pelisoli2023}). The radio emission of both systems is partly compatible with dipolar spin-down \citep{Geng2016,Buckley2017,duPlessis2019}, but also has a significant component resulting from the intrabinary shock with the wind of the companion star. The lack of an optical/infrared counterpart to \gleam\ at the estimated distance of 1.3\,kpc rules out a similar binary system for this source \citep{2022ApJ...940...72R}. However, it does not exclude lower mass companions, or the possibility of a relatively cold isolated MWD.

We recently discovered another ultra-long period radio pulsar, \gpm, which has been active continuously for $> 33$ years \citep{Natasha2023}. \gpm\ has a periodicity of $\sim 21$\,minutes, estimated distance of $\sim5.7$\,kpc and radio luminosity of $10^{27-28}$\,erg\,s$^{-1}$ (peak fluxes of $\sim0.1-10$\,Jy). Its long-duration activity allowed us to infer a stringent constraint on its period derivative of $3.6\times10^{-13}$\,\ss, resulting in strong limits on its $B$-field of $<1.9\times10^{15}$\,G and rotational power $\dot{E}< 2.1\times10^{25}$\,erg\,s$^{-1}$, about 2-3 orders of magnitudes lower than its detected radio luminosity. 

In the following, we study \gleam\ and \gpm\, in the context of the radio emission expected from spin-down of an isolated NS and WD by means of death-line analyses (Sec.~\ref{sec:deathlines}) and population-synthesis simulations (Sec.~\ref{sec:popsyn}). 

\begin{figure*}
\centering
\includegraphics[width=0.8\textwidth]{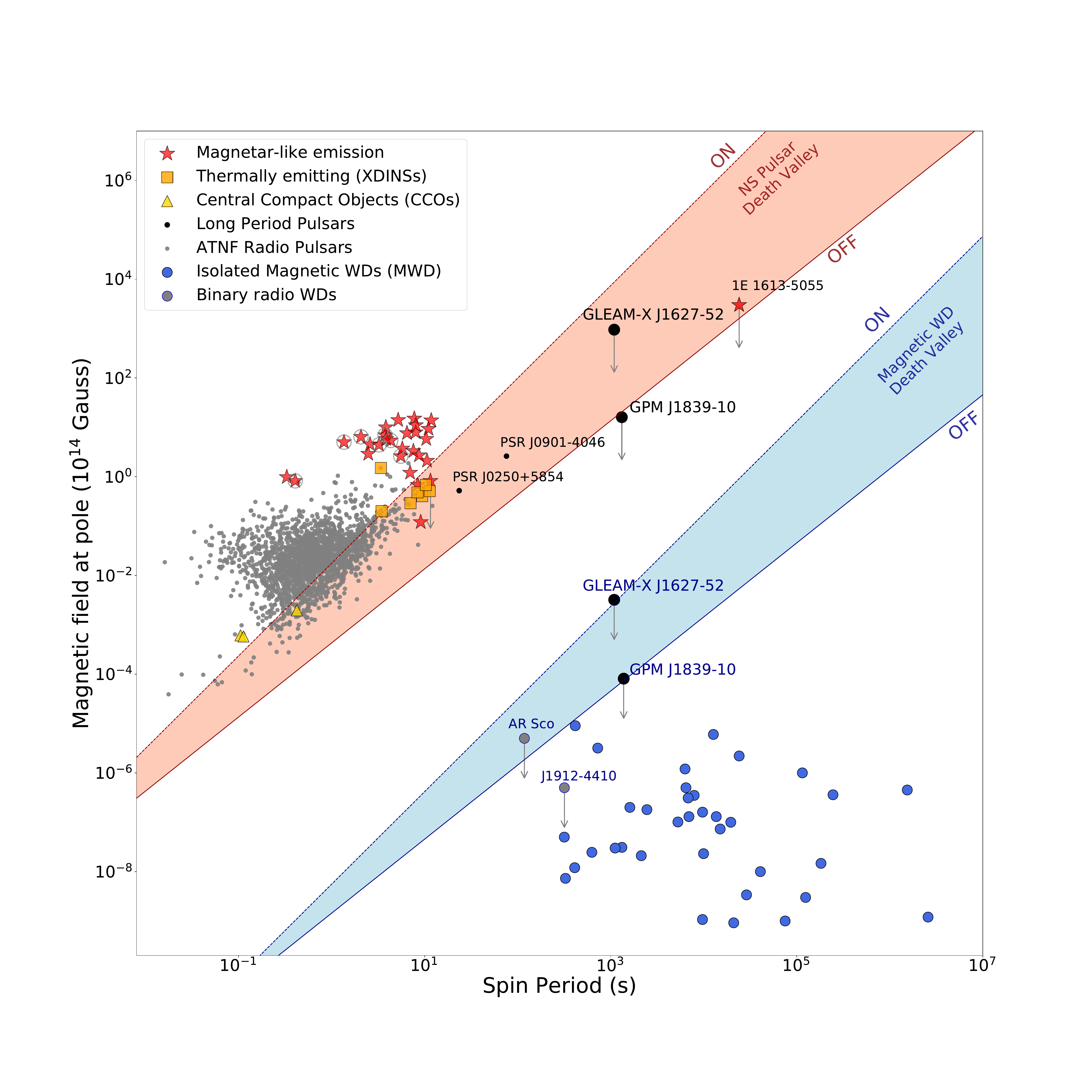}
\caption{Surface dipolar magnetic field, $B$, against spin period, $P$, for observed isolated NSs and magnetic WDs. \gpm{} and \gleam{} are interpreted as isolated NSs or WDs. Arrows represent upper $B$-field limits. We show isolated ATNF radio pulsars \citep{2005AJ....129.1993M} (gray dots), pulsars with magnetar-like X-ray emission (red stars; gray circles highlight the radio magnetars), including the long-period magnetar 1E~161348-5055 \citep{2006Sci...313..814D,2016ApJ...828L..13R,2016MNRAS.463.2394D}, X-ray Dim Isolated NSs (XDINSs; orange squares) and Central Compact Objects (CCOs; gold triangles) \citep{2014ApJS..212....6O, 2018MNRAS.474..961C}. Other long-period radio pulsars are reported as black circles \citep{Tan-etal2018,2022NatAs...6..828C}. Isolated MWDs are represented by blue dots \citep{Ferrario2020,2021Natur.595...39C, Buckley2017}. Gray dots show upper $B$-field limits for the binary WDs AR\,Sco \citep{Buckley2017} and J1912-4410 \citep{Pelisoli2023,Pelisoli2024}. Dashed (solid) lines correspond to theoretical death lines for a pure dipole (extremely twisted multipole) configuration. Red and blue shaded regions indicate NS and WD death valleys, respectively.}
\label{fig:p_B}
\end{figure*}


\section{Death valleys for neutron-star and white-dwarf radio pulsars}
\label{sec:deathlines}

Radio emission from rotating magnetospheres is usually explained as a result of pair production just above the polar caps \citep{1975ApJ...196...51R}. However, for certain limiting periods, magnetic-field strengths and geometries, radio pulsars can no longer produce pairs, and radio emission ceases.
 
The parameter space in the $P-B$ plane (or equivalently $P-\dot{P}$ plane) below which radio emission is quenched is called the ``death valley'' \citep{1993ApJ...402..264C, 2000ApJ...531L.135Z}. This death valley encompasses a large variety of death lines depending on the magnetic-field configuration (e.g., dipolar, multi-polar, twisted), the nature of the seed gamma-ray photons for pair production (i.e., curvature or inverse Compton photons), the pulsar obliquity, the stellar radius and moment of inertia (see \citealt{Suvorov2023} for a death-valley discussion for long-period pulsars). These death-line models have been applied exclusively to NS pulsars because, until very recently, no WD pulsed radio emission had been detected. However, MWDs might not be unlike NS pulsars in generating coherent radio emission through magnetospheric spin-down losses \citep{Zhang2005}, albeit with different stellar radii, masses and magnetic fields.

Figure~\ref{fig:p_B} shows death valleys for NS and WD pulsars as red and blue shaded regions, respectively. Their boundaries are marked by two death-line extremes \citep{1993ApJ...402..264C}, representing the broadest range of $B$-field configurations:\footnote{We note that although these death lines rely on simplifications compared to more recent works \citep[e.g.,][]{2000ApJ...531L.135Z}, our focus is on the extremes of the death valley. All newer models incorporating more detailed physics fall within the shaded region for any reasonable NS or WD parameters.}

\begin{itemize} 
\item a pure dipole with 
\begin{align} \label{eq:deathline_dipole}
B \simeq 2.2\times10^{12} \, {\rm G} \, R_{6}^{-19/8} \left( \frac{P}{1 \, {\rm s}} \right)^{15/8};
\end{align}
\item an extremely twisted, multipolar magnetic field located in a small spot above the polar cap with 
\begin{align} \label{eq:deathline_twisted}
B \simeq 9.2\times10^{10} \,{\rm G} \, \chi^{-1/4} R_{6}^{-2} \left( \frac{P}{1 \, {\rm s}} \right)^{3/2},
\end{align}
where $R_{6} = R/10^{6}$\,cm, and $\chi$ is the ratio between the spot's $B$-field and the dipolar strength (we assume an extreme value of $\chi=10$).
\end{itemize}

For Fig.~\ref{fig:p_B}, we use a fiducial NS radius of $R_\mathrm{NS} = 11$\,km. 
For WDs, we assume $R_\mathrm{WD} = 6000$\,km consistent with the Hamada-Salpeter relation \citep{Hamada1961} and measurements of isolated MWDs \citep{Ferrario2015,Ferrario2020}.


For observed NS pulsars, we derive surface magnetic fields at their poles using $P$ and $\dot{P}$ measurements, employing the classical dipolar loss formula $B = (3 c^3 I P \dot{P}/2 \pi^2 R^6)^{1/2} \simeq 6.4\times10^{19}\sqrt{\dot{P}P}$\,G,
(assuming $M_\mathrm{NS} = 1.4M_{\odot}$). $B$-fields at the poles of isolated WDs are obtained from observations (i.e., Zeeman splitting of spectral lines; \citealt{Ferrario2015,Ferrario2020, 2021Natur.595...39C}). For the radio pulsating WDs AR\,Sco and J1912$-$4410, we estimate upper $B$-field limits assuming the emission to result from dipolar losses \citep{Buckley2017}. Finally, we also show the upper limits on the surface dipolar $B$-fields of the two long-period radio sources \gleam\ and \gpm.



\begin{figure*}
\centering
\includegraphics[width=\textwidth]{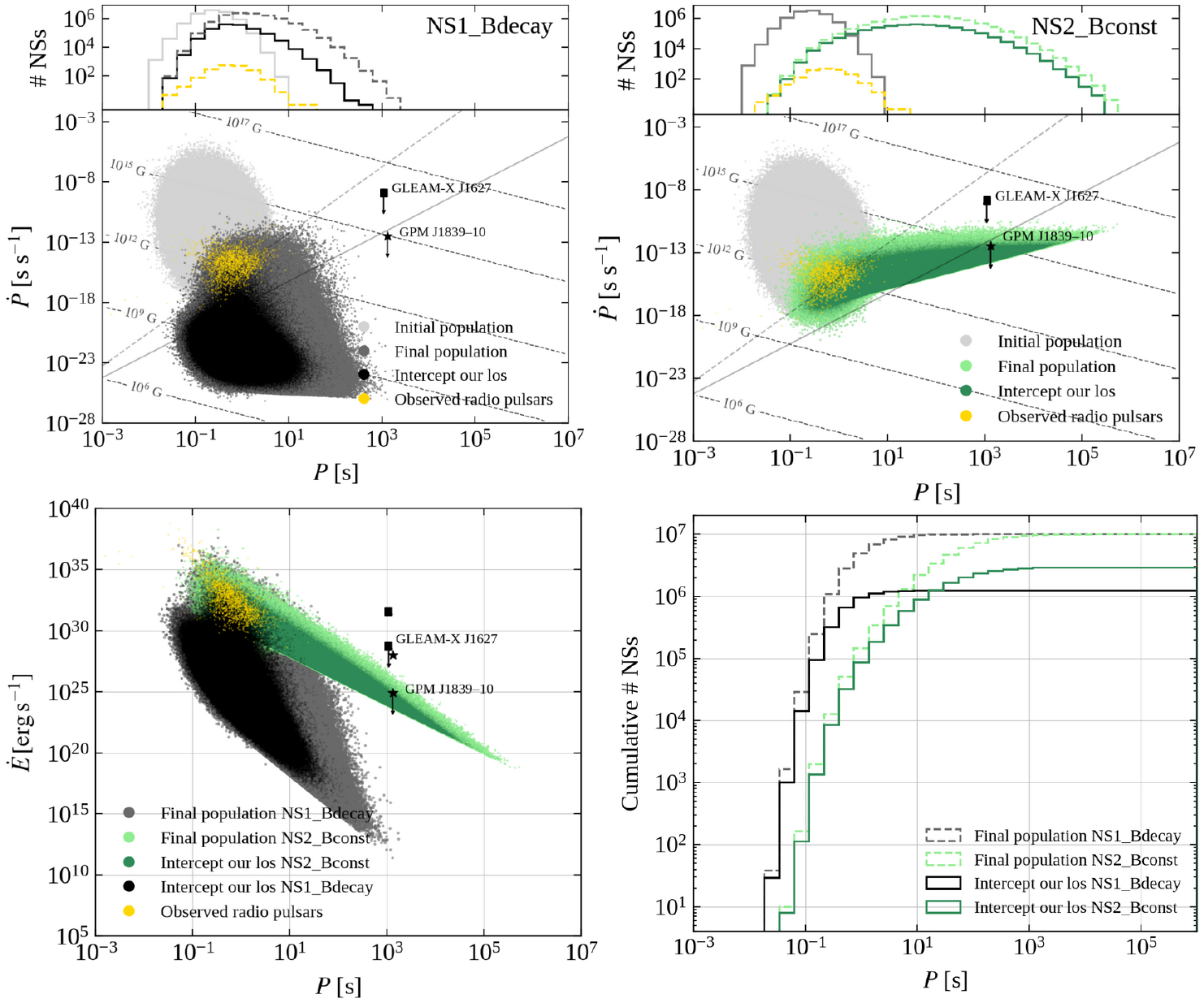}
\caption{Population-synthesis results for models {\tt NS1\_Bdecay} (black) and {\tt NS2\_Bconst} (green). Top panels show $P-\dot{P}$ diagrams for both simulations, respectively. Light gray dots represent initial NS populations, dark gray and light green final populations. The subsets of objects intercepting our line of sight (l.o.s) are shown in black and dark green. Yellow dots are the observed isolated pulsar population from the ATNF Pulsar Catalogue \citep{Manchester2005}. Dotted lines of constant $B$-field are indicated for reference as well as the limits of the death valley and the upper limits for \gleam\ (square) and \gpm\ (star) as in Fig.\,\ref{fig:p_B}. Histograms above the $P-\dot{P}$ diagrams represent the corresponding period distributions. The bottom left panel shows $\dot{E}$ versus $P$ for the evolved populations, where we highlight the radio luminosity (top markers) and upper limits on $\dot{E}$ (bottom arrows) for the two sources. The bottom right panel highlights the cumulative period distributions. }
\label{fig:pop_syn_NS_1}
\end{figure*}

\begin{figure*}
\centering
\includegraphics[width=\textwidth]{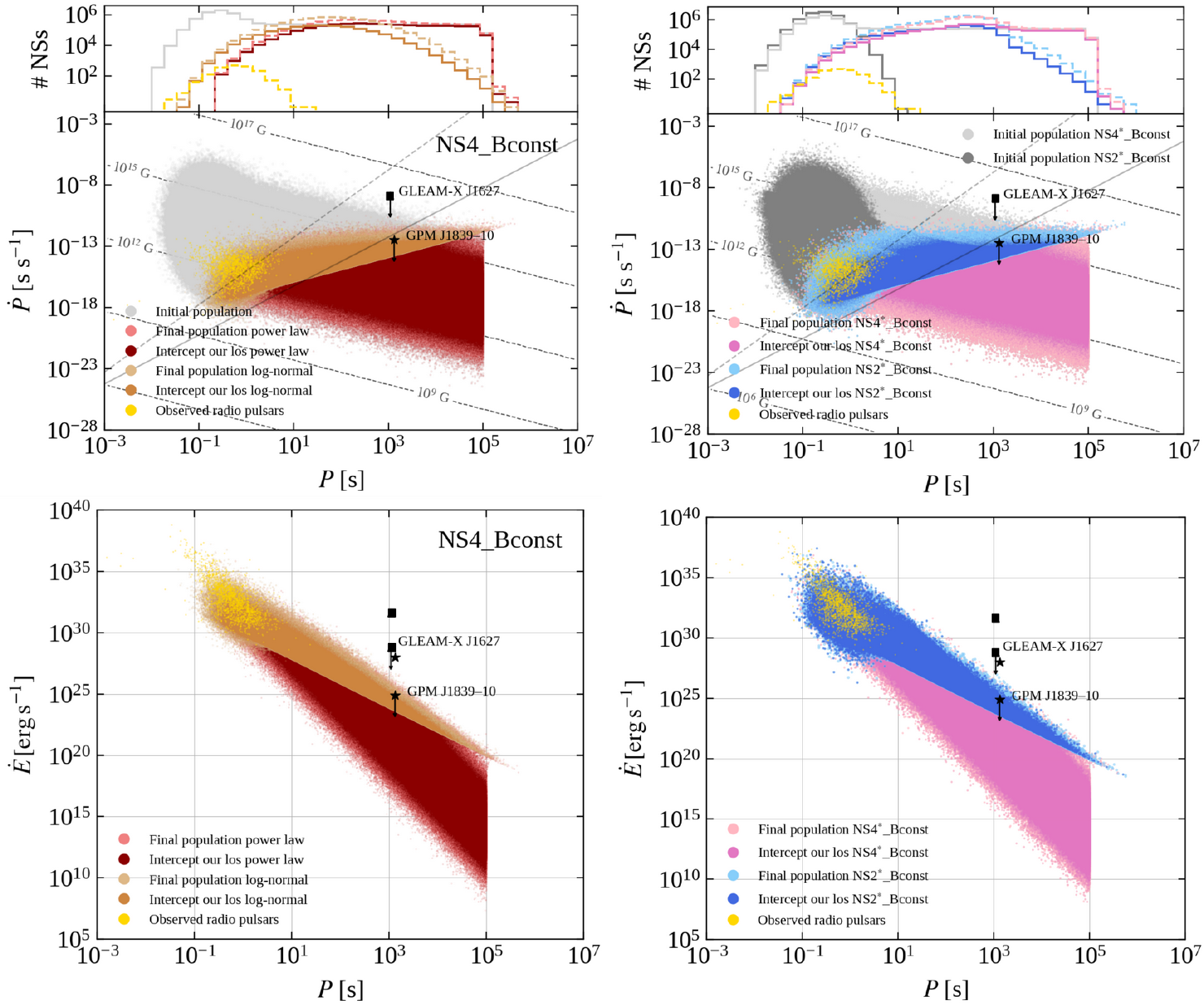}
\caption{Population-synthesis results for models {\tt NS4\_Bconst} (orange/red), {\tt NS2$^*$\_Bconst} (blue), and {\tt NS4$^*$\_Bconst} (pink). Panels, lines, markers and yellow dots are similar to Fig.\,\ref{fig:pop_syn_NS_1}.
In both left panels, evolved objects sampled from the log-normal (power-law) contribution to the initial period distribution are shown in orange (red). Right panels show evolved population of models {\tt NS4$^*$\_Bconst} and {\tt NS2$^*$\_Bconst} based on a bimodal B-field distributions. Across all panels, light shades depict evolved populations, while objects intercepting our lines of sight are shown in dark shades.}
\label{fig:pop_syn_NS_2}
\end{figure*}


\section{Population synthesis for neutron-star and white-dwarf radio pulsars}
\label{sec:popsyn}

We simulate isolated NS and WD populations using the framework of Graber et al. (in prep; see also \cite{Ronchi2021}) with model parameters adjusted for each object type. Initially, we randomly sample the logarithm of the birth periods and magnetic fields from normal distributions, and the inclination angle between the magnetic and the rotational axis from a uniform distribution in spherical coordinates. Assuming that NSs and WDs spin down due to magnetospheric torques, we then evolve their periods, $P$, and inclination angles, $\chi$, over time by solving the coupled differential equations \citep{Philippov2014,Spitkovsky2006}
\begin{align}	\label{eq:magrot}
\dot{P} &= \frac{\pi^2}{c^3}\frac{B^2 R^6}{I P} \left( \kappa_0 + \kappa_1 \sin^2 \chi \right), \\
\dot{\chi} &= -\frac{\pi^2}{c^3}\frac{B^2 R^6}{I P^2} \left( \kappa_2 \sin\chi \cos\chi \right),
\end{align} 
where we assume for simplicity $I = 2/5 M R^2$ and $\kappa_0 \simeq \kappa_1 \simeq \kappa_2 \simeq 1$ for pulsars surrounded by magnetospheres. Finally, we determine the number of stars that point towards the Earth by assuming a random direction for the line of sight and employing a prescription for the aperture of the radio beam. 

\begin{figure*}
\centering
\includegraphics[width=\textwidth]{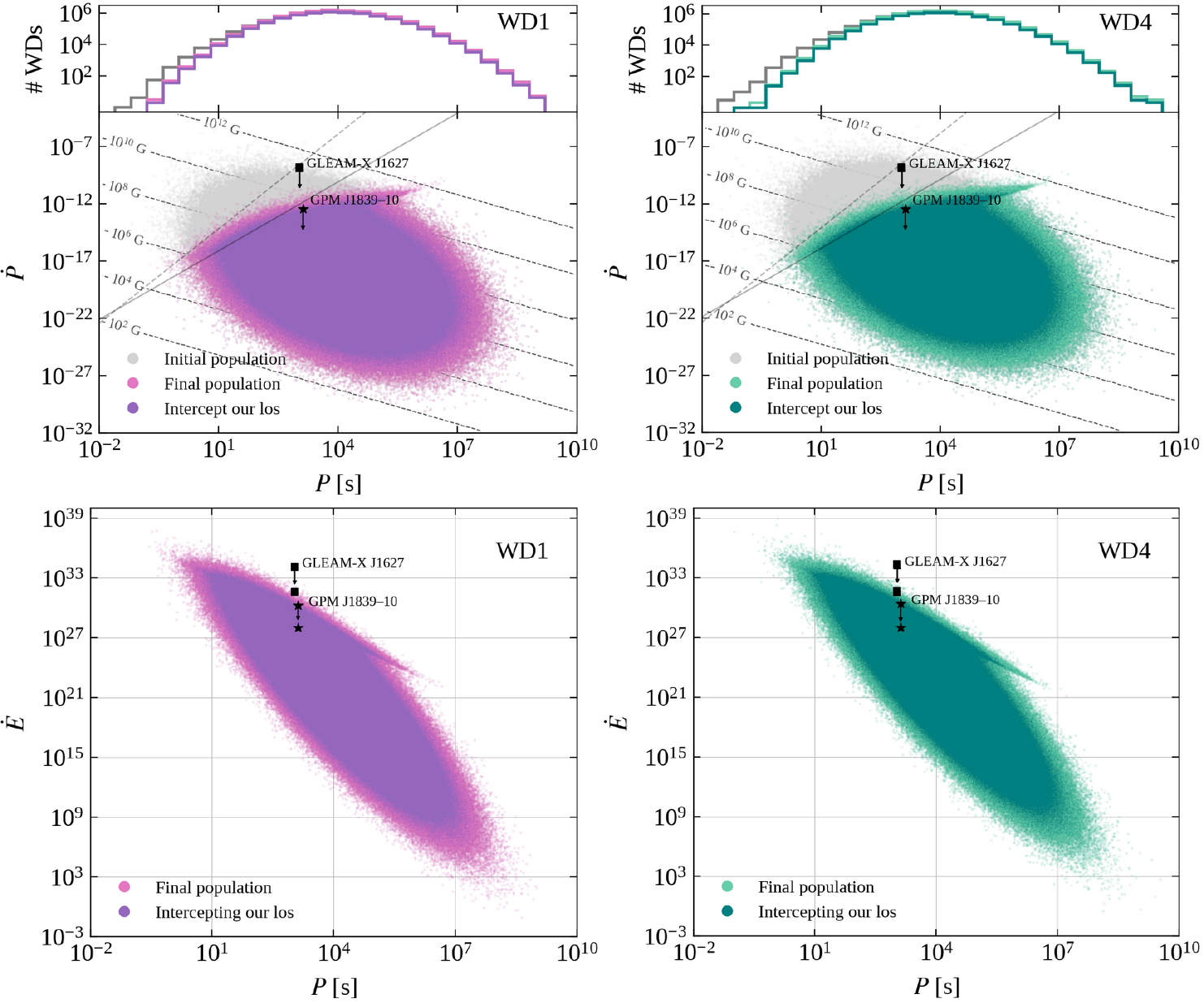}
\caption{Population-synthesis results for models {\tt WD1} (purple) and {\tt WD4} (seagreen). Panels, lines and markers are similar to Fig.\,\ref{fig:pop_syn_NS_1}. Note that the death valley and the upper limits for the two long-period sources refer to the WD case (see Fig.\,\ref{fig:p_B}, Secs.\,\ref{sec:deathlines} and \ref{subsec:popsyn_WD}). Note that these two WD cases have different masses and radii, which is reflected in different $B$-field lines, death-valley and $\dot{E}$ limits.
Gray dots represent initial WD populations, pink and light seagreen final populations. The subsets of objects intercepting our line of sight are shown in purple and dark seagreen. }
\label{fig:pop_syn_WD}
\end{figure*}

To compare the impact of various initial model assumptions on the final spin-period distributions, we carry out the population simulations summarized in Tab.~\ref{tab:results} and Figs.~\ref{fig:pop_syn_NS_1}--\ref{fig:pop_syn_WD}. Specifically, in Tab.\,\ref{tab:results}, to help the reader, we count the objects falling within the period ranges $10-10^2$\,s, $10^2-10^3$\,s, and $10^3-10^5$\,s. We then distinguish objects intercepting our lines of sight and those with $\dot{E} > 10^{27} \, {\rm erg} \, {\rm s}^{-1}$ (see Figs.~\ref{fig:pop_syn_NS_1}--\ref{fig:pop_syn_WD} for the exact $\dot{E}$ and $P$ distributions). The latter limit has no intrinsic meaning, but was chosen as a reference to show how many sources would have sufficient rotational power to support \gpm's radio luminosity.
In Figs.~\ref{fig:pop_syn_NS_1}--\ref{fig:pop_syn_WD}, we also report in the $P-\dot{P}$ plane the two death-line extremes defined in Sec.\,\ref{sec:deathlines}. This is done by making explicit the dependence on the stellar mass and radius and substituting $B = (3 c^3 I P \dot{P}/2 \pi^2 R^6)^{1/2} \simeq 5.7\times10^{19} (M/M_{\odot})^{1/2} R_6^{-2} \sqrt{\dot{P}P}$\,G into Eqs.~\eqref{eq:deathline_dipole} and~\eqref{eq:deathline_twisted}. In this way we obtain:
\begin{align}
\dot{P} \simeq 1.5\times10^{-15} \,{\rm s \, s}^{-1} \, \left( \frac{M}{M_{\odot}}\right)^{-1} R_{6}^{-3/4} \left( \frac{P}{1 \, {\rm s}} \right)^{11/4};
\end{align}
\begin{align}
\dot{P} \simeq 2.6\times10^{-18} \,{\rm s \, s}^{-1} \, \chi^{-1/2} \left( \frac{M}{M_{\odot}}\right)^{-1} \left( \frac{P}{1 \, {\rm s}} \right)^{2};
\end{align}
for a pure dipole or an extremely twisted multipolar field, respectively.
Note that the second death line only depends on the NS or WD mass but not on their radius.

\subsection{NS population synthesis}
\label{subsec:popsyn_NS}

We simulate $10^7$ NSs with random ages sampled from a uniform distribution up to a maximum age of $10^9$ yrs. This translates to a birth rate of one NS per century, consistent with the Galactic core-collapse supernova rate \citep{Rozwadowska2021}. To assign each NS a birth field, we then sample the logarithm of the field (in Gauss) from a normal distribution with mean $\mu_{\log B}=13.25$ and a standard deviation of $\sigma_{\log B} = 0.75$ \citep[see, e.g.,][]{Gullon2014, Gullon2015, Cieslar2020}. Unless stated otherwise, we adopt $M_{\rm NS} = 1.4 \, M_{\odot}$ and $R_{\rm NS} = 11 \,$km. 

We sample the logarithm of the initial period from a normal distribution with mean $\mu_{\log P}$ = -0.6 (corresponding to $0.25$\,s) and standard deviation $\sigma_{\log P} = 0.3$ \citep{Popov2010,Gullon2014, Xu2023}. We further incorporate magnetic-field decay due to Ohmic dissipation and the Hall effect through magneto-thermal evolution curves from \cite{Vigano2013, Vigano2021} and assume a radio beam angular aperture $\propto P^{-1/2}$ \citep{2012hpa..book.....L}. Model {\tt NS1\_Bdecay} serves as a reference with standard population assumptions (Fig.\,\ref{fig:pop_syn_NS_1} top-left). These are typical initial parameters compatible with the current observed pulsar population. However, they do not predict any long-period pulsars.

We continue with investigating more extreme scenarios, focusing first on zero field decay. Strong fields could be maintained over long timescales if electric currents are predominantly present in the NS core \citep[e.g.,][]{Vigano2021}. Consequently, NSs experience a more pronounced spin-down, reaching longer periods.
For model {\tt NS1\_Bconst}, we thus repeat the set-up of {\tt NS1\_Bdecay} but with constant $B$-field 
at the very limit of what is physically viable. However, adding the constant-$B$ assumption is insufficient to slow down the population substantially (see Tab.\,\ref{tab:results}). For subsequent models, we continue with the extreme constant $B$-field case to explore the impact of other assumptions.

In model {\tt NS2\_Bconst}, we also relax the standard beaming assumption, adjusting the radio beam angular aperture to obtain a duty cycle of 20\% in line with observations of \gleam{} and \gpm{}. This results in an increase of the number of pulsars crossing our line of sight (see Fig.\,\ref{fig:pop_syn_NS_1} top-right and bottom panels). For the remaining simulations, we thus maintain this prescription of the beaming unless stated otherwise.

Next, we explore different initial spin-period distributions, mimicking a possible interaction with initial fallback accretion \citep[see, e.g.,][]{Alpar2001, Ertan2009, Tong2016, 2022ApJ...934..184R}. For models {\tt NS3\_Bconst} to {\tt NS6\_Bconst}, we add a power law with an arbitrary cut-off at a period of $10^5$\,s to the aforementioned log-normal distribution of the observed pulsar population. We specifically consider a power law, as the spin-down is likely determined by different fallback accretion rates. Note that the cut-off does not affect our final results, but reflects the maximum spin reachable by fallback accretion: see Figs.\,3--4 of \cite{2022ApJ...934..184R}. We arbitrarily assume that both distributions are equally normalized, sampling 50\% of NSs from either distribution, maintaining a birth rate of 1 NS per century. This prescription is still consistent with the log-normal population resulting in the observed radio pulsars (\cite{Gullon2015}; see also yellow dots in Figs. \ref{fig:pop_syn_NS_1} and \ref{fig:pop_syn_NS_2}). For models {\tt NS3\_Bconst} and {\tt NS4\_Bconst} (see Fig.~\ref{fig:pop_syn_NS_2} left panels), we assume a corresponding power-law index of -3 and -1, respectively. {\tt NS5\_Bconst} investigates a duty cycle of $10\%$, while for {\tt NS6\_Bconst}, we explore the effect of the assumed mass, setting $M_{\rm NS} = 2 M_{\odot}$. 

Since stronger magnetic fields enhance the spin-down, we also investigate the effect of a bimodal $B$-field distribution (four models denoted with an asterisk ($*$) in Tab.~\ref{tab:results}). Besides the log-normal distribution, we consider that 50\% of NSs are formed with a strong field uniformly distributed in $\log B \in [13.5, 14.5]$ following \cite{Gullon2015}. {\tt NS2$^*$\_Bdecay} and {\tt NS2$^*$\_Bconst} consider only the log-normal for the initial period distribution and a decaying and constant magnetic field, respectively, while for {\tt NS4$^*$\_Bdecay} and {\tt NS4$^*$\_Bconst}, we explore the log-normal plus power law for the initial period (see Fig.\ref{fig:pop_syn_NS_2} right panels).

Using the same prescription as \citet{Gullon2014}, we have checked that all population models presented here are consistent with the observed pulsar population. This is mainly due to the low rotational power and long periods of the resulting hidden pulsars.


\subsection{WD population synthesis}
\label{subsec:popsyn_WD}

MWDs spin down slower than NSs due to larger moments of inertia, larger spin periods and lower $B$-fields. 
Moreover, magnetic fields of MWDs do not exhibit relevant magnetic-field decay due to longer Ohmic dissipation timescales \citep[e.g.,][]{Cumming2002} and can be taken as constant \citep{Ferrario2020}. Consequently, current isolated WD periods and magnetic-fields strengths closely reflect those at birth.

To model these birth distributions, we consider a sample of 37 MWDs with reliable spin-period and magnetic-field measurements \citep{Ferrario2020}. We fit Gaussian functions to the distributions of the logarithm of the periods and $B$-fields, deriving a mean of $\mu_{\log P} = 3.94$ and standard deviation of $\sigma_{\log P} = 1.0$, and $\mu_{\log B} = 6.91$ and $\sigma_{\log B} = 1.09$, respectively. For our population synthesis, we then simulate $10^8$ MWDs with ages drawn from a uniform distribution up to a maximum of $10^9\,$yr, consistent with a birth rate of 10 per century derived assuming a Galaxy radius of 20\,kpc and 10\% of the WD being magnetic (see, e.g., \citealt{Holberg2016} but also \citealt{Bagnulo2021} who recently found 22\%), and assign initial $P$ and $B$ values from our fitted distributions. Results of four simulation configurations are summarized in Tab.~\ref{tab:results} and Fig.\,\ref{fig:pop_syn_WD}.

We further assume a WD radio beam angular aperture independent of $P$. For models {\tt WD1} and {\tt WD2}, we adjust our approach to obtain a 20\% and 10\% duty cycle, respectively. For {\tt WD3} and {\tt WD4}, we set the beaming as in {\tt WD1} but vary mass and radius. In particular, using the Hamada-Salpeter mass-radius relation for He WDs \citep{Hamada1961}, we consider $M_{\rm WD} = 1.2 \, M_{\odot}$ and $R_{\rm WD} = 4000$\,km for a high-mass WD in {\tt WD3} and $M_{\rm WD} = 0.6 \, M_{\odot}$ with $R_{\rm WD} = 9000$\,km for a low-mass WD in {\tt WD4}.

\begin{deluxetable*}{l|cc|rrr|rrr}
\tablecaption{Population-synthesis results for isolated NSs and MWDs. \label{tab:results}}
\tabletypesize{\tiny}
\tablecolumns{9}
\tablenum{1}
\tablehead{
\multicolumn{9}{c}{} \\
& & & 
\multicolumn{3}{c}{Total N/$10^3$ [l.o.s.]}  &
\multicolumn{3}{|c}{Total N/$10^3$ with $\dot{E} > 10^{27}$ erg s$^{-1}$ [l.o.s.]}\\
\hline
\colhead{Model} &
\colhead{Initial $P$} &
\colhead{Beaming} &
\colhead{$P = 10^{1-2}$\,s} &
\colhead{$P = 10^{2-3}$\,s} &
\colhead{$P = 10^{3-5}$\,s} &
\colhead{$P = 10^{1-2}$\,s} &
\colhead{$P = 10^{2-3}$\,s} &
\colhead{$P = 10^{3-5}$\,s}}
\startdata
\multicolumn{9}{c}{NEUTRON STARS} \\
\hline
{\tt NS1\_Bdecay}  & Log-N &  pulsars    & 629.51 [9.22] & 6.69 [0.02] & 0.01 [0.00] & 2.49 [0.05] & 0.01 [0.00] & 0.00 [0.00] \\
{\tt NS1\_Bconst} & Log-N  & pulsars    & 4560.95 [31.15] & 3357.42 [3.70] & 574.38 [0.09] & 2797.74 [24.68] & 34.62 [0.06] & 0.00 [0.00]\\
{\tt NS2\_Bconst} & Log-N  & 20\%   &  4564.26 [1360.90] & 3355.02 [799.39] & 573.61 [120.38] & 2799.49 [876.59] & 34.70 [8.75] & 0.00 [0.00]\\
{\tt NS3\_Bconst} & Log-N+PL(-3)  & 20\%    & 4568.02 [1417.86] & 3364.06 [818.03] & 573.68 [122.11] & 2799.46 [915.60] & 34.78 [9.09] & 0.00 [0.00]\\
{\tt NS4\_Bconst} & Log-N+PL(-1)  & 20\%    & 3425.36 [1302.58] &  3344.05 [1341.96] & 2291.68 [1475.93] & 1920.09 [706.18] & 26.38 [8.52] & 0.00 [0.00]\\
{\tt NS5\_Bconst} & Log-N+PL(-1)  & 10\%    & 3425.41 [603.07] & 3345.82 [646.41] & 2289.66 [838.73] &  1920.40 [320.11] & 26.44 [3.62] & 0.00 [0.00]\\
{\tt NS6\_Bconst}\tablenotemark{o} & Log-N+PL(-1) & 20\%   & 3551.79 [1364.31] & 3158.14 [1298.75] & 2200.36 [1453.03] & 2269.94 [837.21] & 37.55 [12.22] & 0.00 [0.00]\\
{\tt NS2\_Bdecay}\tablenotemark{*} & Log-N  & 20\%    & 827.50 [367.95] & 3.37 [0.93] & 0.00 [0.00] & 3.75 [1.74] & 0.00 [0.00] & 0.00 [0.00] \\
{\tt NS2\_Bconst}\tablenotemark{*} & Log-N  & 20\%    & 2621.41 [765.23] & 5849.46 [1342.22] & 774.43 [168.48] & 1656.60 [504.92] & 77.12 [19.16] & 0.00 [0.00] \\
{\tt NS4\_Bdecay}\tablenotemark{*} & Log-N+PL(-1)  & 20\%    & 1511.96 [977.49] & 895.09 [665.54] & 1786.98 [1330.37] & 3.92 [2.27] & 0.01 [0.00] & 0.00 [0.00] \\
{\tt NS4\_Bconst}\tablenotemark{*} & Log-N+PL(-1)  & 20\%    & 1942.87 [717.60] & 5077.88 [1733.97] & 2508.41 [1565.64] & 1136.06 [406.37] & 58.33 [18.82] & 0.00 [0.00] \\  
\hline
\multicolumn{9}{c}{WHITE DWARFS} \\
\hline
{\tt WD1} & Log-N &  20\%  & 1929.55 [1406.56] & 14485.85 [10649.08] & 69050.05 [51234.73] & 1823.72 [1328.54] & 7956.49 [5803.80] & 3880.60 [2729.91] \\
{\tt WD2} &  Log-N  & 10\%  & 1928.76 [817.16] & 14484.90 [6225.08] & 69042.10 [30034.14] & 1822.90 [771.63] & 7955.59 [3386.88] & 3883.30 [1577.85] \\
{\tt WD3}\tablenotemark{o} &  Log-N   & 20\%   & 2147.21 [1576.08] & 14705.23 [10871.34] & 68592.09 [51016.48] & 1894.94 [1389.56] & 5687.17 [4172.56] & 1664.71 [1176.52] \\
{\tt WD4}\tablenotemark{o} &  Log-N & 20\%  & 1579.74 [1141.63] & 13837.27 [10076.93] & 70003.37 [51617.09] & 1542.90 [1114.49] & 9672.34 [6989.46] & 7872.76 [5454.37] \\
\enddata
\tablecomments{All models were evolved for $10^{9}$yr assuming a birth rate of 1 NS and 10 MWD per century. Unless specified otherwise, NSs have 1.4$M_{\odot}$ and 11\,km radii, while MWDs have 1$M_{\odot}$ and 6000\,km radii. Numbers in parentheses denote the assumed PL slope. Numbers in brackets are the sub-sample of simulated neutron stars that cross our line of sight [l.o.s.] for their respective beaming. Note that all reported numbers are in units of $10^{3}$, hence being integers.} 
\tablenotetext{*}{Bimodal magnetic field distribution (see text).}
\tablenotetext{o}{{\tt NS6\_Bconst} assumes a 2$M_{\odot}$ NS mass, {\tt WD3} a mass of 1.2$M_{\odot}$ and 4000\,km radius, and {\tt WD4} a mass of 0.6$M_{\odot}$ and 9000\,km radius.}
\end{deluxetable*}


\section{Discussion and conclusion}
\label{sec:conclusion}

Wide-field radio interferometers have begun to revolutionize our understanding of the transient radio sky. Until recently, coherent, polarized and periodic radio emission was characteristic of NS pulsars with periods $\lesssim 20$\,s, a period range attributed to magnetic-field decay and a resistive crust \citep{Pons2013}. However, in the last year, two ultra-long period systems, \gleam\, and \gpm\, \citep{2022Natur.601..526H,Natasha2023}, and the slow pulsar PSR J0901-4046 \citep{2022NatAs...6..828C} were discovered. 


In this work, we study long-period pulsars in the rotating NS and WD dipole scenario, one of the most likely interpretations given their coherent and highly polarized emission, and perform population synthesis. 

1. While the classical scenario for NS pulsar radio emission based on magnetospheric pair production can in principle accommodate \gleam, it cannot account for \gpm\, as the source sits below even the most extreme death line (Fig.~\ref{fig:p_B}). However, note that both objects have radio luminosities exceeding their $\dot{E}$s by 2-3 orders of magnitudes \citep{Natasha2023}. Hence, the emission scenario is necessarily more complex than for normal radio pulsars (possibly resembling radio magnetars; see Fig.\,6 in \citealt{2022ApJ...940...72R}).

2. Figure~\ref{fig:p_B} also highlights that a similar mechanism in MWDs could in principle contribute to the radio emission of \gleam, AR\,Sco and J1912$-$4410. However, \gpm's bright radio emission cannot be easily reconciled even in the isolated MWD case. Note that interactions with a companion's wind can enhance the radio emission (as for AR\,Sco and J1912$-$4410; see also \citealt{Geng2016}). For \gleam\ optical and IR observations could rule out main sequence companion stars \citep{2022ApJ...940...72R}, but for \gpm\, a similar constrain was not possible given the larger distance \citep{Natasha2023}. 

3. Our NS population synthesis models (Tab.~\ref{tab:results}) show that a large population of long-period radio emitters cannot be easily explained as NS pulsars. Neither standard population assumptions nor the most extreme scenarios invoking zero field decay (Fig.~\ref{fig:pop_syn_NS_1}), initial slow-down via fallback accretion, 20\% duty cycles or stronger birth fields (Fig.~\ref{fig:pop_syn_NS_2}) result in sufficiently energetic NS pulsars with periods $>1000$\,s pointing towards the Earth (irrespective of mass). A difference by a factor of a few in the NS birth rate does not alter this conclusion.

4. On the other hand, WD population synthesis highlights that long-period MWD are more common than NS pulsars, with lower masses and larger radii leading to enhanced spin down and more slow rotators (Tab.~\ref{tab:results}). 
However, Fig.~\ref{fig:p_B} shows that the known sample of isolated MWDs are not expected to emit coherent radio emission via standard pair production, being all below the most extreme death line. 

In conclusion, the classical particle acceleration mechanism for rotating dipoles fails to provide a satisfactory explanation for the radio emission of \gpm\ in either the NS or WD scenario. 
In contrast, all observed isolated MWDs with measured $B$-fields fall below the most extreme death lines, possibly explaining their radio non-detection. The radio emission observed from the binary WDs AR\,Sco and J1912$-$4410 might be enhanced by the presence of their companion star within the WD pulsar lightcylinder. Deep optical and IR observations ruled out main sequence stars for \gleam\ \citep{2022ApJ...940...72R}, but deeper observations are needed to exclude any binarity. On the other hand, \gpm's limits \citep{Natasha2023} cannot provide strong constraints given its larger distance.

Moreover, in the NS scenario, we do not expect a large population of ultra-long period pulsars under any (physically motivated or extreme) assumptions. While many more slow WD pulsars might be expected, however we still lack a mechanism to explain the bright radio emission. Therefore, if \gleam\, and \gpm\, are confirmed as isolated NS or WD pulsars, this would call for a revision of our understanding of radio emission from dipolar magnetospheres. Corroborating the NS scenario would further require a significant reexamination of our understanding of initial NS parameters (birth rates, magnetic-field distribution, etc.) at the population level.


N.R., C.P.A, M.R., V.G. and F.C.Z. are supported by the ERC via the Consolidator Grant ``MAGNESIA'' (No. 817661), and by the program Unidad de Excelencia Mar\'ia de Maeztu CEX2020-001058-M. We acknowledges partial support from grant SGR2021-01269 (PI: Graber). C.P.A. and M.R.'s work has been carried out within the framework of the doctoral program in Physics of the Universitat Autonoma de Barcelona. N.H.W. is the recipient of an ARC Future Fellowship (No. FT190100231). F.C.Z. is also supported by a Ram\'on y Cajal Fellowship (RYC2021-030888-I). Dd.M. acknowledges financial support from ASI and INAF. We would like to thank A.~Kawka, J.~Isern, A.~Harding, R.~Turolla, D. Vigan\'o and J.~A. Pons for discussions. We also acknowledge the anonymous referee for the useful comments and suggestions that improved the manuscript.


\begin{thebibliography}{}
\expandafter\ifx\csname natexlab\endcsname\relax\def\natexlab#1{#1}\fi
\providecommand{\url}[1]{\href{#1}{#1}}

\bibitem[{{Alpar} {et~al.}(2001){Alpar}, {Ankay}, \& {Yazgan}}]{Alpar2001}
{Alpar}, M.~A., {Ankay}, A., \& {Yazgan}, E. 2001, \apjl, 557, L61

\bibitem[{{Bagnulo} \& {Landstreet}(2021)}]{Bagnulo2021}
{Bagnulo}, S., \& {Landstreet}, J.~D. 2021, \mnras, 507, 5902

\bibitem[{{Bhattacharya} \& {van den Heuvel}(1991)}]{B&H1991}
{Bhattacharya}, D., \& {van den Heuvel}, E.~P.~J. 1991, \physrep, 203, 1

\bibitem[{{Buckley} {et~al.}(2017){Buckley}, {Meintjes}, {Potter}, {Marsh}, \&
  {G{\"a}nsicke}}]{Buckley2017}
{Buckley}, D.~A.~H., {Meintjes}, P.~J., {Potter}, S.~B., {Marsh}, T.~R., \&
  {G{\"a}nsicke}, B.~T. 2017, Nature Astronomy, 1, 0029

\bibitem[{{Caiazzo} {et~al.}(2021){Caiazzo}, {Burdge}, {Fuller}, {Heyl},
  {Kulkarni}, {Prince}, {Richer}, {Schwab}, {Andreoni}, {Bellm}, {Drake},
  {Duev}, {Graham}, {Helou}, {Mahabal}, {Masci}, {Smith}, \&
  {Soumagnac}}]{2021Natur.595...39C}
{Caiazzo}, I., {Burdge}, K.~B., {Fuller}, J., {et~al.} 2021, \nat, 595, 39

\bibitem[{{Caleb} {et~al.}(2022){Caleb}, {Heywood}, {Rajwade}, {Malenta},
  {Stappers}, {Barr}, {Chen}, {Morello}, {Sanidas}, {van den Eijnden},
  {Kramer}, {Buckley}, {Brink}, {Motta}, {Woudt}, {Weltevrede}, {Jankowski},
  {Surnis}, {Buchner}, {Bezuidenhout}, {Driessen}, \&
  {Fender}}]{2022NatAs...6..828C}
{Caleb}, M., {Heywood}, I., {Rajwade}, K., {et~al.} 2022, Nature Astronomy, 6,
  828

\bibitem[{{Chatterjee} {et~al.}(2000){Chatterjee}, {Hernquist}, \&
  {Narayan}}]{Chatterjee2000}
{Chatterjee}, P., {Hernquist}, L., \& {Narayan}, R. 2000, \apj, 534, 373

\bibitem[{{Chen} \& {Ruderman}(1993)}]{1993ApJ...402..264C}
{Chen}, K., \& {Ruderman}, M. 1993, \apj, 402, 264

\bibitem[{{Cie{\'s}lar} {et~al.}(2020){Cie{\'s}lar}, {Bulik}, \&
  {Os{\l}owski}}]{Cieslar2020}
{Cie{\'s}lar}, M., {Bulik}, T., \& {Os{\l}owski}, S. 2020, \mnras, 492, 4043

\bibitem[{{Coti Zelati} {et~al.}(2018){Coti Zelati}, {Rea}, {Pons}, {Campana},
  \& {Esposito}}]{2018MNRAS.474..961C}
{Coti Zelati}, F., {Rea}, N., {Pons}, J.~A., {Campana}, S., \& {Esposito}, P.
  2018, \mnras, 474, 961

\bibitem[{{Cumming}(2002)}]{Cumming2002}
{Cumming}, A. 2002, \mnras, 333, 589

\bibitem[{{D'A{\`\i}} {et~al.}(2016){D'A{\`\i}}, {Evans}, {Burrows}, {Kuin},
  {Kann}, {Campana}, {Maselli}, {Romano}, {Cusumano}, {La Parola}, {Barthelmy},
  {Beardmore}, {Cenko}, {De Pasquale}, {Gehrels}, {Greiner}, {Kennea}, {Klose},
  {Melandri}, {Nousek}, {Osborne}, {Palmer}, {Sbarufatti}, {Schady}, {Siegel},
  {Tagliaferri}, {Yates}, \& {Zane}}]{2016MNRAS.463.2394D}
{D'A{\`\i}}, A., {Evans}, P.~A., {Burrows}, D.~N., {et~al.} 2016, \mnras, 463,
  2394

\bibitem[{{De Luca} {et~al.}(2006){De Luca}, {Caraveo}, {Mereghetti}, {Tiengo},
  \& {Bignami}}]{2006Sci...313..814D}
{De Luca}, A., {Caraveo}, P.~A., {Mereghetti}, S., {Tiengo}, A., \& {Bignami},
  G.~F. 2006, Science, 313, 814

\bibitem[{{du Plessis} {et~al.}(2019){du Plessis}, {Wadiasingh}, {Venter}, \&
  {Harding}}]{duPlessis2019}
{du Plessis}, L., {Wadiasingh}, Z., {Venter}, C., \& {Harding}, A.~K. 2019,
  \apj, 887, 44

\bibitem[{{Ertan} {et~al.}(2009){Ertan}, {Ek{\c{s}}i}, {Erkut}, \&
  {Alpar}}]{Ertan2009}
{Ertan}, {\"U}., {Ek{\c{s}}i}, K.~Y., {Erkut}, M.~H., \& {Alpar}, M.~A. 2009,
  \apj, 702, 1309

\bibitem[{{Ferrario} {et~al.}(2015){Ferrario}, {de Martino}, \&
  {G{\"a}nsicke}}]{Ferrario2015}
{Ferrario}, L., {de Martino}, D., \& {G{\"a}nsicke}, B.~T. 2015, \ssr, 191, 111

\bibitem[{{Ferrario} {et~al.}(2020){Ferrario}, {Wickramasinghe}, \&
  {Kawka}}]{Ferrario2020}
{Ferrario}, L., {Wickramasinghe}, D., \& {Kawka}, A. 2020, Advances in Space
  Research, 66, 1025

\bibitem[{{Ferrario} \& {Wickramasinghe}(2005)}]{Ferrario2005}
{Ferrario}, L., \& {Wickramasinghe}, D.~T. 2005, \mnras, 356, 615

\bibitem[{{Geng} {et~al.}(2016){Geng}, {Zhang}, \& {Huang}}]{Geng2016}
{Geng}, J.-J., {Zhang}, B., \& {Huang}, Y.-F. 2016, \apjl, 831, L10

\bibitem[{{Gull{\'o}n} {et~al.}(2014){Gull{\'o}n}, {Miralles}, {Vigan{\`o}}, \&
  {Pons}}]{Gullon2014}
{Gull{\'o}n}, M., {Miralles}, J.~A., {Vigan{\`o}}, D., \& {Pons}, J.~A. 2014,
  \mnras, 443, 1891

\bibitem[{{Gull{\'o}n} {et~al.}(2015){Gull{\'o}n}, {Pons}, {Miralles},
  {Vigan{\`o}}, {Rea}, \& {Perna}}]{Gullon2015}
{Gull{\'o}n}, M., {Pons}, J.~A., {Miralles}, J.~A., {et~al.} 2015, \mnras, 454,
  615

\bibitem[{{Hamada} \& {Salpeter}(1961)}]{Hamada1961}
{Hamada}, T., \& {Salpeter}, E.~E. 1961, \apj, 134, 683

\bibitem[{{Holberg} {et~al.}(2016){Holberg}, {Oswalt}, {Sion}, \&
  {McCook}}]{Holberg2016}
{Holberg}, J.~B., {Oswalt}, T.~D., {Sion}, E.~M., \& {McCook}, G.~P. 2016,
  \mnras, 462, 2295

\bibitem[{{Hurley-Walker} {et~al.}(2023){Hurley-Walker}, {Rea}, {McSweeney}, \&
  {et}}]{Natasha2023}
{Hurley-Walker}, N., {Rea}, N., {McSweeney}, S.~J., \& {et}, A. 2023, \nat, ?,
  ?

\bibitem[{{Hurley-Walker} {et~al.}(2022){Hurley-Walker}, {Zhang}, {Bahramian},
  {McSweeney}, {O'Doherty}, {Hancock}, {Morgan}, {Anderson}, {Heald}, \&
  {Galvin}}]{2022Natur.601..526H}
{Hurley-Walker}, N., {Zhang}, X., {Bahramian}, A., {et~al.} 2022, \nat, 601,
  526

\bibitem[{{Lorimer} \& {Kramer}(2012)}]{2012hpa..book.....L}
{Lorimer}, D.~R., \& {Kramer}, M. 2012, {Handbook of Pulsar Astronomy}
  (Cambridge University Press)

\bibitem[{{Manchester} {et~al.}(2005{\natexlab{a}}){Manchester}, {Hobbs},
  {Teoh}, \& {Hobbs}}]{2005AJ....129.1993M}
{Manchester}, R.~N., {Hobbs}, G.~B., {Teoh}, A., \& {Hobbs}, M.
  2005{\natexlab{a}}, \aj, 129, 1993

\bibitem[{{Manchester} {et~al.}(2005{\natexlab{b}}){Manchester}, {Hobbs},
  {Teoh}, \& {Hobbs}}]{Manchester2005}
---. 2005{\natexlab{b}}, \aj, 129, 1993

\bibitem[{{Marsh} {et~al.}(2016){Marsh}, {G{\"a}nsicke}, {H{\"u}mmerich},
  {Hambsch}, {Bernhard}, {Lloyd}, {Breedt}, {Stanway}, {Steeghs}, {Parsons},
  {Toloza}, {Schreiber}, {Jonker}, {van Roestel}, {Kupfer}, {Pala}, {Dhillon},
  {Hardy}, {Littlefair}, {Aungwerojwit}, {Arjyotha}, {Koester}, {Bochinski},
  {Haswell}, {Frank}, \& {Wheatley}}]{2016Natur.537..374M}
{Marsh}, T.~R., {G{\"a}nsicke}, B.~T., {H{\"u}mmerich}, S., {et~al.} 2016,
  \nat, 537, 374

\bibitem[{{Olausen} \& {Kaspi}(2014)}]{2014ApJS..212....6O}
{Olausen}, S.~A., \& {Kaspi}, V.~M. 2014, \apjs, 212, 6

\bibitem[{{Pelisoli} {et~al.}(2023){Pelisoli}, {Marsh}, {Buckley}, {Heywood},
  {Potter}, {Schwope}, {Brink}, {Standke}, {Woudt}, {Parsons}, {Green},
  {Kepler}, {Munday}, {Romero}, {Breedt}, {Brown}, {Dhillon}, {Dyer}, {Kerry},
  {Littlefair}, {Sahman}, \& {Wild}}]{Pelisoli2023}
{Pelisoli}, I., {Marsh}, T.~R., {Buckley}, D. A.~H., {et~al.} 2023, Nature
  Astronomy, arXiv:2306.09272

\bibitem[{{Pelisoli} {et~al.}(2024){Pelisoli}, {Sahu}, {Lyutikov}, {Barkov},
  {G{\"a}nsicke}, {Brink}, {Buckley}, {Potter}, {Schwope}, \&
  {Ram{\'\i}rez}}]{Pelisoli2024}
{Pelisoli}, I., {Sahu}, S., {Lyutikov}, M., {et~al.} 2024, \mnras, 527, 3826

\bibitem[{{Philippov} {et~al.}(2014){Philippov}, {Tchekhovskoy}, \&
  {Li}}]{Philippov2014}
{Philippov}, A., {Tchekhovskoy}, A., \& {Li}, J.~G. 2014, \mnras, 441, 1879

\bibitem[{{Pons} {et~al.}(2013){Pons}, {Vigan{\`o}}, \& {Rea}}]{Pons2013}
{Pons}, J.~A., {Vigan{\`o}}, D., \& {Rea}, N. 2013, Nature Physics, 9, 431

\bibitem[{{Popov} {et~al.}(2010){Popov}, {Pons}, {Miralles}, {Boldin}, \&
  {Posselt}}]{Popov2010}
{Popov}, S.~B., {Pons}, J.~A., {Miralles}, J.~A., {Boldin}, P.~A., \&
  {Posselt}, B. 2010, \mnras, 401, 2675

\bibitem[{{Rea} {et~al.}(2016){Rea}, {Borghese}, {Esposito}, {Coti Zelati},
  {Bachetti}, {Israel}, \& {De Luca}}]{2016ApJ...828L..13R}
{Rea}, N., {Borghese}, A., {Esposito}, P., {et~al.} 2016, \apjl, 828, L13

\bibitem[{{Rea} {et~al.}(2022){Rea}, {Coti Zelati}, {Dehman}, {Hurley-Walker},
  {de Martino}, {Bahramian}, {Buckley}, {Brink}, {Kawka}, {Pons}, {Vigan{\`o}},
  {Graber}, {Ronchi}, {Pardo Araujo}, {Borghese}, {Parent}, \&
  {Galvin}}]{2022ApJ...940...72R}
{Rea}, N., {Coti Zelati}, F., {Dehman}, C., {et~al.} 2022, \apj, 940, 72

\bibitem[{{Ronchi} {et~al.}(2021){Ronchi}, {Graber}, {Garcia-Garcia}, {Rea}, \&
  {Pons}}]{Ronchi2021}
{Ronchi}, M., {Graber}, V., {Garcia-Garcia}, A., {Rea}, N., \& {Pons}, J.~A.
  2021, \apj, 916, 100

\bibitem[{{Ronchi} {et~al.}(2022){Ronchi}, {Rea}, {Graber}, \&
  {Hurley-Walker}}]{2022ApJ...934..184R}
{Ronchi}, M., {Rea}, N., {Graber}, V., \& {Hurley-Walker}, N. 2022, \apj, 934,
  184

\bibitem[{{Rozwadowska} {et~al.}(2021){Rozwadowska}, {Vissani}, \&
  {Cappellaro}}]{Rozwadowska2021}
{Rozwadowska}, K., {Vissani}, F., \& {Cappellaro}, E. 2021, \na, 83, 101498

\bibitem[{{Ruderman} \& {Sutherland}(1975)}]{1975ApJ...196...51R}
{Ruderman}, M.~A., \& {Sutherland}, P.~G. 1975, \apj, 196, 51

\bibitem[{{Spitkovsky}(2006)}]{Spitkovsky2006}
{Spitkovsky}, A. 2006, \apjl, 648, L51

\bibitem[{{Suvorov} \& {Melatos}(2023)}]{Suvorov2023}
{Suvorov}, A.~G., \& {Melatos}, A. 2023, \mnras, 520, 1590

\bibitem[{{Tan} {et~al.}(2018){Tan}, {Bassa}, {Cooper}, {Dijkema}, {Esposito},
  {Hessels}, {Kondratiev}, {Kramer}, {Michilli}, {Sanidas}, {Shimwell},
  {Stappers}, {van Leeuwen}, {Cognard}, {Grie{\ss}meier}, {Karastergiou},
  {Keane}, {Sobey}, \& {Weltevrede}}]{Tan-etal2018}
{Tan}, C.~M., {Bassa}, C.~G., {Cooper}, S., {et~al.} 2018, \apj, 866, 54

\bibitem[{{Tauris}(2012)}]{Tauris2012}
{Tauris}, T.~M. 2012, Science, 335, 561

\bibitem[{{Tong} {et~al.}(2016){Tong}, {Wang}, {Liu}, \& {Xu}}]{Tong2016}
{Tong}, H., {Wang}, W., {Liu}, X.~W., \& {Xu}, R.~X. 2016, \apj, 833, 265

\bibitem[{{Vigan{\`o}} {et~al.}(2021){Vigan{\`o}}, {Garcia-Garcia}, {Pons},
  {Dehman}, \& {Graber}}]{Vigano2021}
{Vigan{\`o}}, D., {Garcia-Garcia}, A., {Pons}, J.~A., {Dehman}, C., \&
  {Graber}, V. 2021, Computer Physics Communications, 265, 108001

\bibitem[{{Vigan{\`o}} {et~al.}(2013{\natexlab{a}}){Vigan{\`o}}, {Rea}, {Pons},
  {Perna}, {Aguilera}, \& {Miralles}}]{2013MNRAS.434..123V}
{Vigan{\`o}}, D., {Rea}, N., {Pons}, J.~A., {et~al.} 2013{\natexlab{a}},
  \mnras, 434, 123

\bibitem[{{Vigan{\`o}} {et~al.}(2013{\natexlab{b}}){Vigan{\`o}}, {Rea}, {Pons},
  {Perna}, {Aguilera}, \& {Miralles}}]{Vigano2013}
---. 2013{\natexlab{b}}, \mnras, 434, 123

\bibitem[{{Xu} {et~al.}(2023){Xu}, {Yang}, {Mao}, {Xu}, {Li}, \&
  {Liu}}]{Xu2023}
{Xu}, K., {Yang}, H.-R., {Mao}, Y.-H., {et~al.} 2023, \apj, 947, 76

\bibitem[{{Zhang} \& {Gil}(2005)}]{Zhang2005}
{Zhang}, B., \& {Gil}, J. 2005, \apjl, 631, L143

\bibitem[{{Zhang} {et~al.}(2000){Zhang}, {Harding}, \&
  {Muslimov}}]{2000ApJ...531L.135Z}
{Zhang}, B., {Harding}, A.~K., \& {Muslimov}, A.~G. 2000, \apjl, 531, L135

\end{thebibliography}

\end{document}